\documentclass[letterpaper]{article} 
\usepackage{aaai2026}
\usepackage{times}  
\usepackage{helvet}  
\usepackage{courier}  
\usepackage[hyphens]{url}  
\usepackage{graphicx} 
\usepackage{natbib}  
\usepackage{caption} 
\usepackage{algorithm}
\usepackage{algorithmic}
\usepackage{amsmath}
\usepackage{multirow}
\usepackage{makecell}
\usepackage{newfloat}
\usepackage{listings}
\usepackage{pifont}

\usepackage{xcolor}

\DeclareCaptionStyle{ruled}{labelfont=normalfont,labelsep=colon,strut=off} 
\floatstyle{ruled}
\newfloat{listing}{tb}{lst}{}
\floatname{listing}{Listing}
\title{Removing Box-Free Watermarks for Image-to-Image Models via Query-Based Reverse Engineering}
\author{
    Haonan An\textsuperscript{\rm 1},
    Guang Hua\textsuperscript{\rm 2},
    Hangcheng Cao\textsuperscript{\rm 1},
    Zhengru Fang\textsuperscript{\rm 1},
    Guowen Xu\textsuperscript{\rm 3},
    Susanto Rahardja\textsuperscript{\rm 2},
    Yuguang Fang\textsuperscript{\rm 1}
}
\affiliations{
    \textsuperscript{\rm 1}Department of Computer Science, City University of Hong Kong, Hong Kong\\
    \textsuperscript{\rm 2}Infocomm Technology and Engineering Cluster, Singapore Institute of Technology, Singapore 828608\\
    \textsuperscript{\rm 3}School of Computer Science and Engineering, University of Electronic Science and Technology of China
}

\usepackage{bibentry}

\nocopyright
\begin{document}

\maketitle

\begin{abstract}
The intellectual property of deep generative networks (GNets) can be protected using a cascaded hiding network (HNet) which embeds watermarks (or marks) into GNet outputs, known as box-free watermarking. Although both GNet and HNet are encapsulated in a black box (called operation network, or ONet), with only the generated and marked outputs from HNet being released to end users and deemed secure, in this paper, we reveal an overlooked vulnerability in such systems. Specifically, we show that the hidden GNet outputs can still be reliably estimated via query-based reverse engineering, leaking the generated and unmarked images, despite the attacker's limited knowledge of the system. Our first attempt is to reverse-engineer an inverse model for HNet under the stringent black-box condition, for which we propose to exploit the query process with specially curated input images. While effective, this method yields unsatisfactory image quality. To improve this, we subsequently propose an alternative method leveraging the equivalent additive property of box-free model watermarking and reverse-engineering a forward surrogate model of HNet, with better image quality preservation. Extensive experimental results on image processing and image generation tasks demonstrate that both attacks achieve impressive watermark removal success rates ($100\%$) while also maintaining excellent image quality (reaching the highest PSNR of $34.69$ dB), substantially outperforming existing attacks, highlighting the urgent need for robust defensive strategies to mitigate the identified vulnerability in box-free model watermarking.
\end{abstract}

\section{Introduction}
Recently, deep neural networks (DNNs), especially generative networks (GNets), have demonstrated their powerful ability to handle various tasks, surpassing previous state-of-the-art techniques. However, the resources required to train such models, whether in terms of time, money, or labor, are immense. For example, the widely used generative DNN application GPT-4 requires more than 24,000 GPUs for training \cite{LIU2023100017}, resulting in significant expenses. Therefore, it is imperative to safeguard these assets from intellectual property infringement.

Previous efforts in safeguarding the intellectual property of DNN models aim at two primary goals: 1) Ownership verification and 2) Model stealing tracing, also known as surrogate attack tracing. Both goals can be accomplished through watermarking techniques. The watermarking process embeds marks, e.g., encoded identity information, into the to-be-protected model or its outputs. Subsequently, a verification process can extract the embedded information to verify the ownership. In the case of model stealing tracing, model watermarking has shown its effectiveness in retaining the embedded marks in surrogate models, which can then be extracted through the verification process.

\begin{figure}[!t]
  \centering
  \includegraphics[width=0.45\textwidth]{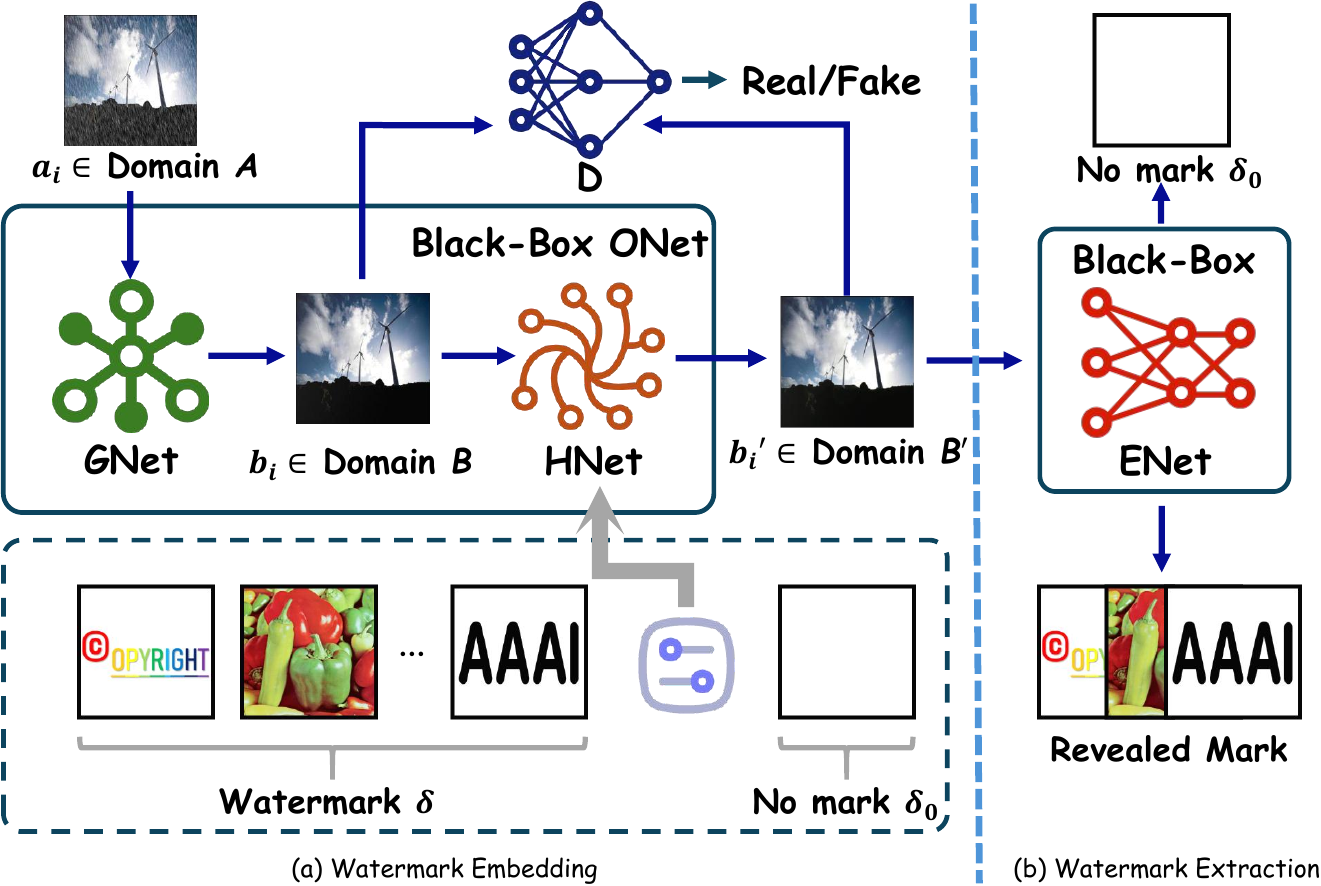}
  \caption{Flowchart of the victim model considered in this paper, where image denoising is used as an example.
  }
  \label{fig:victim_model}
\end{figure}
Among all categories of model watermarking, box-free watermarking stands out due to its ability to handle high-entropy outputs (e.g., images) and its flexibility in extracting watermarks solely from protected models’ outputs. In this subdomain, Zhang et al. \cite{zhang2020model} proposed the first box-free framework for image processing models. Although existing box-free methods have shown robustness against common image processing, surrogate attacks, and applicability to other generative tasks \cite{wu2020watermarking, zhang2021deep, huang2023can, zhang2024robust, an2025decoder}, in this paper, we reveal an overlooked vulnerability that query-based reverse engineering can leak the generated and unmarked images, enabling watermark removal.


We build on the prior works \cite{wu2020watermarking, zhang2020model, zhang2021deep, zhang2024robust} as depicted in Figure \ref{fig:victim_model}. Considering image denoising as an example, the noisy image, denoted by $a_i \in A$, is processed by GNet, whose output, denoised image $b_i \in B$, is not released. Instead, $b_i$ is further processed by the hiding network (HNet) for watermark embedding, resulting in a processed and marked image, $b_i' \in B'$, which is the actual output of the model. Notably, GNet and HNet are encapsulated as a black-box called operation network (ONet). The extraction network (ENet) takes $b_i'$ as input to retrieve the embedded watermark, while an all-white image indicates the absence of watermarks.

Our attacks are grounded in a key observation that if an attacker can craft queries to circumvent the functionality of GNet, they may extract HNet's information to infer the watermark embedding mechanism, thereby enabling watermark removal against $b_i'$. Based on this, we first present a simple and intuitive method to reverse-engineer an inversion network of the HNet, denoted by $\text{HNet}^{-1}$, which is a watermark removal operator to $b_i'$. However, such an inversion-based removal attack resulted in limited image quality, which motivates us to propose the second attack. We notice that the watermark embedding process in the victim model in Figure~\ref{fig:victim_model} is additive in nature, which can be modeled by $b_i' = b_i + \delta_{bi}'$, where $\delta_{bi}'$ is the equivalent representation of the to-be-embedded watermark $\delta$. Built on this, we demonstrate how the additive property can be exploited to estimate the private $b_i$. Both attacks are evaluated under the realistic and challenging black-box condition, where the attacker only has access to the victim model's inputs ($a_i$) and outputs ($b_i'$). Our contributions are as follows:
\begin{itemize}
    \item  We discover, via query-based reverse engineering, that encapsulating GNet and HNet into a black box is still insecure in protecting GNet and its outputs.
    \item We show that HNet can be inverted via specially crafted queries, which leads to watermark removal and leaking of GNet outputs.
    \item We further propose an improved remover leveraging on the additive equivalence of the watermark, which achieves better image quality.
    \item We discuss a query screener to be deployed at the API to defend against the proposed attacks.
    \item We conduct extensive experiments on image processing and generation tasks to demonstrate the effectiveness of our proposed methods.
\end{itemize}
A comprehensive review of related work is provided in the supplementary material.

\begin{table}[!t]
\renewcommand{\arraystretch}{1.0}
\centering
\caption{List of notations.}
\label{tab:notation} 
\setlength{\tabcolsep}{4pt}
{\begin{tabular}{c|c}
\hline
\hline
Notation  & Definition \\
\hline
$\rm{GNet}$ & Generative network \\
$\rm{HNet}$ &  Hiding network\\
$\rm{ONet}$ &  Operation network ($\rm{GNet} + \rm{HNet}$)\\
$\rm{ENet}$ &  Extraction network\\
$\rm{SNet}$ &  Surrogate network of $\rm{HNet}$\\
$\rm{D}$ &  Discriminator\\
$i, j$ & Sample index, $i \ne j$\\
$\delta_0$ & All-white image (no watermark) \\
$\delta$ & Watermark image \\
$\delta'_{ai}$ & Latent representation of $\delta$ for $a_i$\\
$\delta'_{bi}$ & Latent representation of $\delta$ for $b_i$\\
$a_i \in A$ & To-be-processed image (input of ${\rm{GNet}}$) \\
$b_i \in B$ & Processed unmarked (by ${\rm{GNet}}$) image\\
$b_i' \in B'$ & Processed marked (by ${\rm{ONet}}$) image\\
$\hat{b}_i \in \hat{B}$ & Watermark removed $b_i'$ \\
$e_i \in E$ & Generic unmarked image ($B \subset E$)\\
$b_j \in B$ & Processed unmarked (not by ${\rm{GNet}}$) image\\
$b_j' \in B'$ & $b_j$ marked by ${\rm{HNet}}$ \\
\hline
\hline
\end{tabular}}
\end{table}

\section{Problem Formulation}
\subsection{Box-free Model Watermarking Basics}
\label{sec:foundation}
The existing shared box-free model watermarking workflow is depicted in Figure \ref{fig:victim_model}, and the related notations are summarized in Table \ref{tab:notation}. The workflow is modeled as
\begin{itemize}
    \item Image operation: $b_i = {{\rm{GNet}}(a_i)}$,
    \item Watermark embedding: $b_i'={\rm{HNet}}({\rm{Concat}}(b_i, \delta))$,
    \item Watermark extraction: $\hat{\delta}={\rm{ENet}}(b_i')$,
\end{itemize}
where GNet can be deraining, image generation, or an arbitrary image-to-image model, ${\rm{Concat()}}$ is the channel-wise concatenation operation, and the extraction yields an estimated watermark $\hat{\delta}$ which is supposed to be $\delta$. Note that ENet can also take an unmarked $b_i$ or an arbitrary image, $e_i$, as input, and the output is expected to be $\delta_0$. 
To protect GNet, it is pretrained and frozen, while the defender jointly trains HNet, D, and ENet, to minimize the combined loss 
\begin{equation}
    \mathcal{L}_\text{Joint} = \beta_1 \mathcal{L}_\text{Fidelity} + \beta_2 \mathcal{L}_\text{Mark} + \beta_3 \mathcal{L}_\text{Adv},
\end{equation}
where
\begin{align}
& \mathcal{L}_\text{Fidelity} = \sum\limits_i{\rm{MSE}}\left( b'_i,b_i \right),\\
& \mathcal{L}_\text{Mark} = \sum\limits_i\left[{\rm{MSE}}(\hat{\delta}, \delta) + {\rm{MSE}}\left( {\rm{ENet}}(e_i), \delta_0 \right)\right],\\
& \mathcal{L}_\text{Adv} = \sum\limits_i\left[\log({\rm{D}}(b_i)) + \log(1 - {\rm{D}}(b_i'))\right].
\end{align}
In the above framework, $\mathcal{L}_\text{Fidelity}$ ensures the marked image is visually indistinguishable from the processed unmarked image, $\mathcal{L}_\text{Mark}$ ensures successful watermark extraction from marked images $b_i'$ as well as successful null extraction (all-white output) from unmarked images $e_i$ which can be the processed unmarked $b_i$ or any images not related to GNet. Additionally, the adversarial loss $\mathcal{L}_\text{Adv}$ improves the embedding quality so the discriminator cannot distinguish marked and unmarked images. Other loss functions, e.g., consistency loss \cite{zhang2020model} and perceptual loss \cite{johnson2016perceptual}, can be further added to improve the performance.

\subsection{Threat Model}
We consider a stringent yet practical threat model in this paper, in which GNet and HNet are encapsulated together as a black-box API, referred to as operation network (ONet), and deployed as a cloud service. End users can thus only have their query image $a_i$ and observe the processed and marked image $b_i'$. Meanwhile, ENet is assumed to be operated by an authorized party for watermark extraction and verification, and it is inaccessible to end users including attackers.
The goal of the attack is to remove the watermark embedded in ONet output without significantly degrading image quality, allowing for further attacks like watermark-free surrogate model training. That is to say, ideally, the attacker aims to restore from the observed $b_i'$ to $b_i$, removing the watermark while preserving the effect of GNet. The generic attacking model is thus given by 
\begin{equation}\label{eq:generic}
    {\rm{Remove}}(a_i, b_i') = b_i.
\end{equation}
Note that ${\rm{ENet}}(b_i)=\delta_0$ corresponds to the ideal null extraction, while the removal attack may also be considered successful if ${\rm{ENet}}({\rm{Remove}}(a_i, b_i'))$ significantly differs from $\delta$. The successful estimation of $b_i$ from the only observable $a_i$ and $b_i'$ reveals the vulnerability from the black-box ONet, which is analyzed in the next section. 

\section{Proposed Methods}
In this section, we first reveal the overlooked vulnerability in box-free model watermarking. Building on this, we propose query-based reverse engineering attacks that efficiently remove the embedded watermark. The first attack trains an inversion network of HNet to remove the embedded watermark with simple and intuitive insight. However, we observe the unsatisfactory image quality preservation with this method. To address this, we propose the second attack based on our observation of the additive equivalence property in box-free model watermarking, which demonstrates significantly improved performance in maintaining output image quality. Last, we summarize the proposed attacks and discuss a potential defensive mechanism.

\subsection{Vulnerability Analysis}
From an attacker's perspective, GNet and HNet appear tightly coupled in ONet because the intermediate output $b_i$ is unobtainable. Therefore, breaching this coupling to extract information from either model becomes a critical step in compromising the watermarking system. 

Since ONet is deployed as a commercial cloud service, its functionality is typically known to both users and potential attackers. Our attacks are based on a key observation: if the attacker can construct inputs $b_j \in B$, $j \ne i$ that satisfy approximate identity transformation under GNet, i.e., 
\begin{equation}
    \label{eq:identity_trans}
    \text{GNet}(b_j) \approx b_j,
\end{equation}
then GNet can be effectively bypassed, thereby exposing the watermarking mechanism implemented by HNet. For example, when GNet is instantiated as a deraining model, $b_j$ can be images captured in rain-free environments. When GNet functions as an image generation model, such as Stable Diffusion \cite{rombach2022high_stable_diffusion}, $b_j$ can be masked images where only a small patch is provided for operation. Building on this insight, we develop two attack strategies described in the following sections. 

\subsection{First Attack: Inversion of HNet}
The query process with $b_j$ as inputs can be expressed as
\begin{equation}
    \label{eq:bypass_gnet}
    {\rm{ONet}}(b_j, \delta) = {\rm{HNet}}( {\rm{GNet}}(b_j), \delta) \approx {\rm{HNet}}(b_j, \delta) = b_j',
\end{equation}
where the approximation is based on (\ref{eq:identity_trans}). With a curated set of $b_j$ and $b_j'$, we are able to train a surrogate model of the inversion of HNet, denoted as $\text{HNet}^{-1}$, to efficiently conduct watermark removal by minimizing following removal loss
\begin{equation}
    \label{eq:inversion_HNet}
    \mathcal{L}_{\text{Removal}} = \sum\limits_j{\rm{MSE}} \left({\rm{HNet^{-1}}}(b_j'), b_j\right).
\end{equation}
The number of queries needed to achieve a well-trained $\text{HNet}^{-1}$ is within acceptable range, as shown in our experiments. However, an inferior performance in image quality for $\hat{b}_i = \text{HNet}^{-1}(b_i')$ is observed when compared to $b_i$, which could negatively impact the performance of further attacks. One hypothesis is that the inherent high non-linearity of the DNN model renders the training of an optimal inverse network exceedingly challenging, particularly under conditions where the distribution of training data varies and the architecture of the HNet remains unknown to the attacker. In the next section, we introduce our second attack, which trains a surrogate model of HNet rather than its inversion. This method exploits the additive equivalent property observed in box-free model watermarking to directly estimate $b_i$, resulting in substantial image quality improvement.

\subsection{Second Attack: Forward HNet}

\begin{figure}[!t]
  \centering
  \includegraphics[width=1\columnwidth]{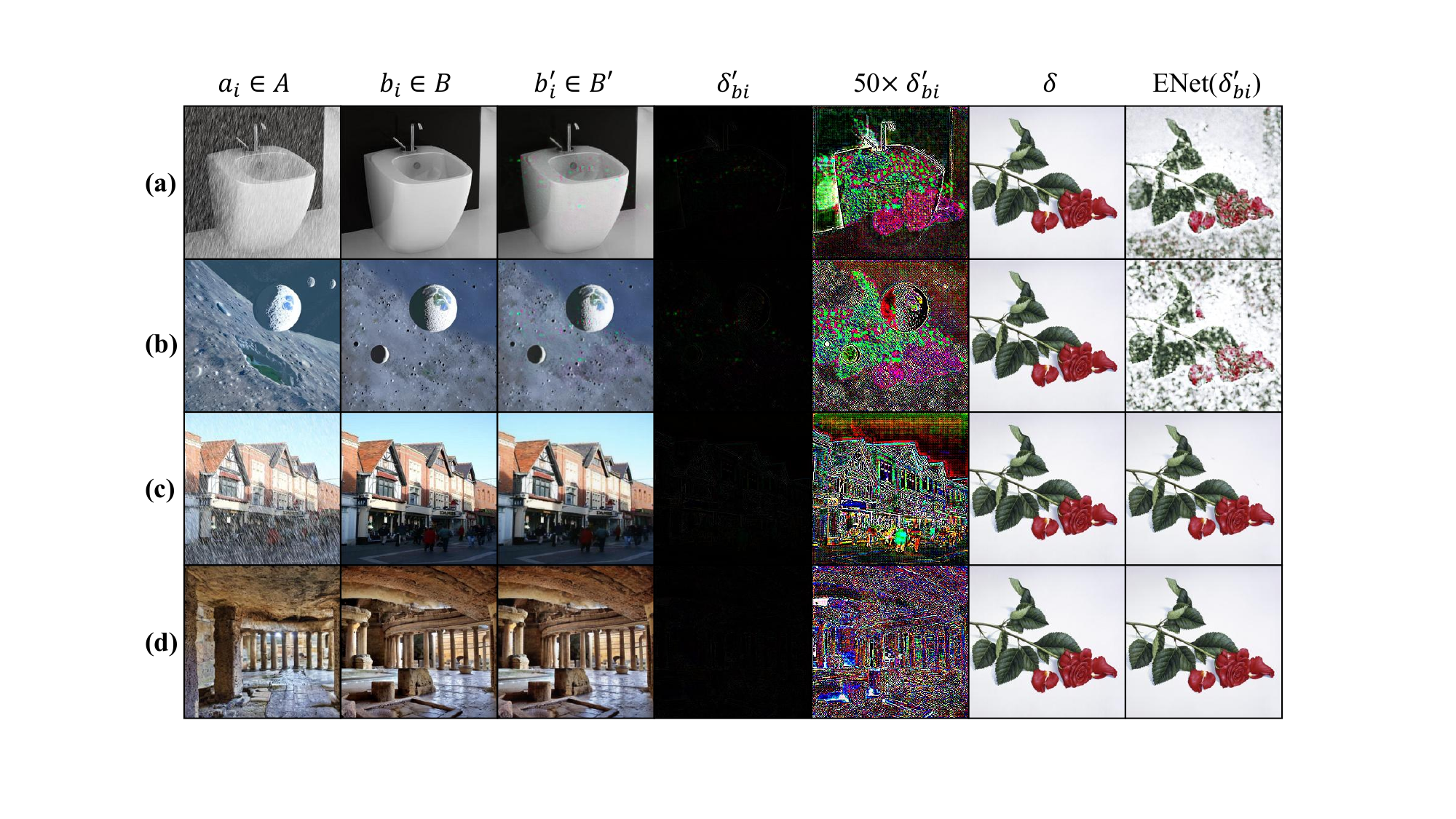}
  \caption{Demonstration of watermark extraction using the difference between $b_i'$ and $b_i$ as the input to ENet to verify the additive property. (a) Deraining  with box-free model watermarking in \cite{wu2020watermarking}. (b) Text-image-to-image-based image editing with box-free model watermarking in \cite{wu2020watermarking} and prompt ``Emphasized planet appearance''. (c) Deraining with box-free model watermarking in \cite{zhang2024robust}. (d) Text-image-to-image-based image editing with box-free model watermarking in \cite{zhang2024robust} and prompt ``Roman bath ruins''.}
  \label{fig:additive_equivalent}
\end{figure}

\begin{figure*}[!t]
  \centering
  \includegraphics[width=1.9\columnwidth]{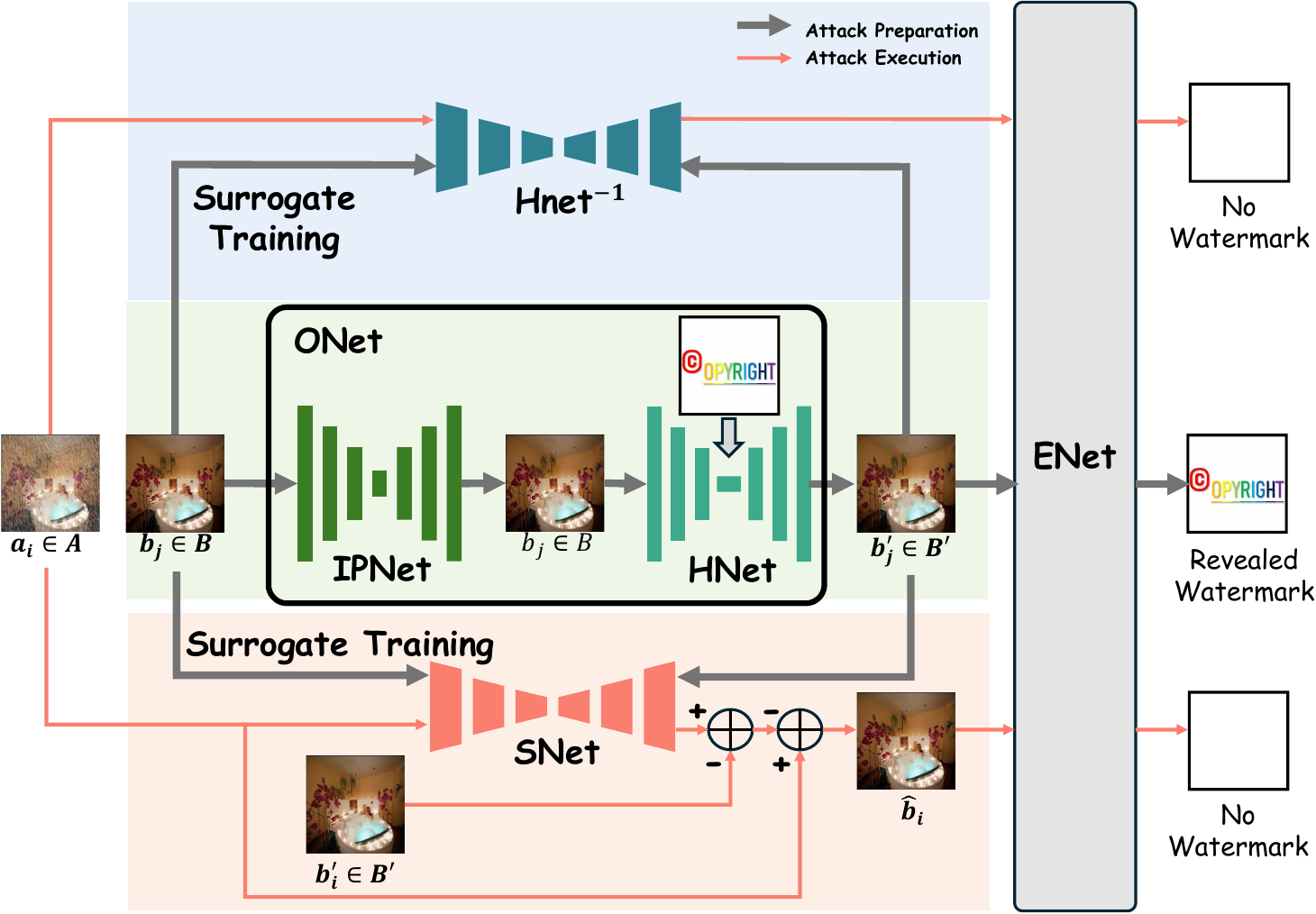}
  \caption{Workflow of victim model and the proposed attacks. Images in Domain $B$, which are watermark-free, can be estimated from all known quantities, showing watermark-free (all-white) outputs from ENet.}
  \label{fig:general_architecture}
\end{figure*}

\subsubsection{The Additive Equivalent}
\label{sec:additive analysis}
We note that the watermark embedding process is a nonlinear high-dimensional mapping from the unmarked $b_i$ to the marked $b_i'$ governed by HNet, but it can also be equivalently expressed as an additive form in which the additive expression of watermark is the residual between $b_i'$ and $b_i$, i.e.,
\begin{equation}\label{eq:additive0}
    b_i' = b_i + \delta'_{bi},
\end{equation}
where $\delta'_{bi}$ is the additive equivalent of the watermark $\delta$ embedded into $b_i$, and in our considered models, $\delta'_{bi}$ is the latent representation of $\delta$. The justification for (\ref{eq:additive0}) stems from the first-order Taylor expansion of $\rm{HNet}(b_i, \delta)$ around origin point of $\delta$
\begin{align}
\label{eq:taylor_derivation}
&{\rm{HNet}}(b_i, \delta) \approx 
{\rm{HNet}}(b_i, 0) + 
\nabla_{\delta} {\rm{HNet}}(b_i, 0) \cdot \delta \notag \\
&\Rightarrow b_i' - b_i \approx  \nabla_{\delta} {\rm{HNet}}(b_i, 0) \cdot \delta = \delta_{bi}', 
\end{align}
where ${\rm{Concat()}}$ is omitted for brevity in (\ref{eq:taylor_derivation}). It reveals that $\delta'_{bi}$ can be approximated as a certain transformation employed to watermark $\delta$ by HNet, i.e., the latent representation of $\delta$. It then holds that
\begin{align}\label{eq:additive}
    \delta'_{bi} & = b_i' - b_i \notag\\
    & = {\rm{HNet}}({\rm{Concat}}(b_i, \delta)) - b_i.
\end{align}

\noindent We can verify the above analysis by extracting the watermark using the residual as input to ENet and check if it holds that
\begin{equation}\label{eq:verify}
    {\rm{ENet}}(\delta'_{bi}) = \delta.
\end{equation}
The verification results are shown in Figure \ref{fig:additive_equivalent}. It can be seen from the $\delta'_{bi}$ column, or more clearly observed in its $50\times$ amplified visualization, that the residual keeps the texture information about the image content but has most semantic information lost. However, it is successfully verified that the watermark can still be reliably extracted from $\delta'_{bi}$. This indicates that ENet can not only extract the watermark from marked images, i.e., ${\rm{ENet}}(b_i') = \delta$, but also restore from the watermark's latent representation to its original image form as shown in (\ref{eq:verify}), in the absence of attacks.

\subsubsection{Estimate of $b_i$}

We now demonstrate that the additive nature of box-free watermarking can be exploited to expose the private information about $b_i$ which should not be released in the black-box threat model. Replace $b_i$ in (\ref{eq:additive}) by $a_i$ and according to (\ref{eq:additive0}), we have
\begin{align}
    {\rm{HNet}}({\rm{Concat}}(a_i, \delta)) & = a_i + \delta_{ai}'\notag\\
    & \approx a_i + \delta_{bi}'\notag\\
    & = a_i + b_i' - b_i,
\end{align}
which yields
\begin{equation}\label{eq:leakage4}
    b_i = a_i - ({\rm{HNet}}({\rm{Concat}}(a_i, \delta)) - b_i'),
\end{equation}

\noindent where the rationale of approximation $\delta_{ai}' \approx \delta_{bi}'$ is based on $\rm{ENet}(\delta'_{ai}) \approx \rm{ENet}(\delta'_{bi})$ and is verified in quantitative experiment. The portion $({\rm{HNet}}({\rm{Concat}}(a_i, \delta)) - b_i')$ is the to-be-processed component for GNet, e.g., the noise component in denoising, the bone component in deboning, and transformation residual in image generation. Note that in (\ref{eq:leakage4}), both $a_i$ and $b_i'$ are available to the attacker, while ${\rm{HNet}}({\rm{Concat}}(a_i, \delta))$ is unknown since it is encapsulated in a black box. However, the attacker can curate special data to query ONet, similar to the first attack, and the outputs of these queries can reveal the underlying functionality of HNet, enabling the creation of a surrogate hiding network that approximates forward HNet, as illustrated in the next subsection.

\subsubsection{Attack Process}
We perform a query-based reverse engineering similar to the first attack by curating $b_j$ to bypass GNet. With the curated $b_j$ and $b_j'$ pairs, instead of training an inverse of HNet, we train a surrogate model denoted by SNet that approximates HNet, establishing the guess mapping from processed but unmarked images to processed and watermarked images by minimizing the following simple loss function
\begin{equation}
    \label{eq:surrogate_loss}
    \mathcal{L}_{\text{Surrogate}} = \sum\limits_j {\rm{MSE}}\left({\rm{SNet}}(b_j), b_j'\right).
\end{equation}
It is important to note here that despite SNet approximates HNet, it does not concatenate the input with a mark but instead directly process $b_j$. The rationale lies in that $b_j'$ inherently contains the mark $\delta$ embedded by HNet. Upon convergence, SNet grabs the functional essence of the black-box protected HNet. Therefore, we can replace the unknown component ${\rm{HNet}}({\rm{Concat}}(a_i, \delta))$ in (\ref{eq:leakage4}) by its approximate SNet, yielding
\begin{equation}
    \label{eq:attack equation}
    \hat{b}_i = a_i - ({\rm{SNet}}(a_i) - b_i'),
\end{equation}
obtaining the estimation of the processed but unmarked (equivalently, watermark-removed) image $b_i$, with all components known to the attacker. Note that (\ref{eq:attack equation}) is our proposed realization of the generic removal attack in (\ref{eq:generic}).

\subsection{Summary and Further Discussion}
To summarize, both attacks operate in two steps: attack preparation and attack execution, as illustrated in Figure \ref{fig:general_architecture}. 
We note that these attacks essentially rely on the requirement for identity transformation to bypass GNet. As a countermeasure, we propose implementing API detection to evaluate the similarity between the input $a_i$ and the output $b_i$ of GNet. A practical implementation involves computing the Euclidean distance of $a_i$ and $b_i$. If the distance falls below a predefined threshold, the system directly returns $a_i$ with warning.

\section{Experiments}
We demonstrate the effectiveness of our proposed approaches by attacking two state-of-the-art box-free model watermarking methods \cite{wu2020watermarking, zhang2024robust} for the tasks of image deraining and image generation, respectively, with the latter focusing on the text-image-to-image-based image editing using Stable Diffusion \cite{rombach2022high_stable_diffusion}. For the ease of notation, \cite{wu2020watermarking} is referred to as $\text{V}_\text{Wu}$ and \cite{zhang2024robust} as $\text{V}_\text{Zhang}$. Notably, $\text{V}_\text{Zhang}$ is the promoted version of \cite{zhang2021deep}, which alleviates the vulnerability against normal image augmentation attacks. However, since our attack method does not involve any image augmentation operations, and the watermark embedding and extraction processes in \cite{zhang2021deep} and ``$\text{V}_\text{Zhang}$'' are identical, we treat both as the same victim and avoid redundant discussion. 

\subsection{Settings}

\subsubsection{Dataset}
Following victims, The PASCAL VOC dataset \cite{everingham2010pascal} is used for image deraining task. It is composed of $12,000$ images from Domain $A$ with raindrop noise, which is generated by algorithm \cite{zhang2018density}, and $12,000$ derained images from Domain $B$. We equally divide the dataset into two parts, each containing $6,000$ noised and $6,000$ denoised images, to serve as training data for the victim models and both attacks, respectively. For image generation, we randomly generate $12,000$ images (served as $a_i$) by Stable Diffusion \cite{rombach2022high_stable_diffusion} and also split them evenly for training the victim model and attacks. In addtion, all the images in both datasets are $256 \times 256$ RGB images.

\subsubsection{Metric}
We evaluate the quality of watermark-removed images by two commonly used metrics, i.e., peak signal-to-noise ratio (PSNR) and multi-scale structural similarity index (MS-SSIM) \cite{wang2003multiscale}, respectively. The watermark removal success rate of our proposed attack is defined as
\begin{equation}
    {\rm{SR}}_\text{Remove} = 1 - {\rm{SR}}_\text{Extract},
\end{equation}
where ${\rm{SR}}_\text{Extract}$ is the rate of successful watermark extractions, and a single extraction is successful if the normalized correlation coefficient between the ENet output and the ground-truth watermark $\delta$ is greater than $0.96$.

\subsection{Qualitative Results}
\begin{figure}[!t]

  \centering
  \includegraphics[width=1\columnwidth]{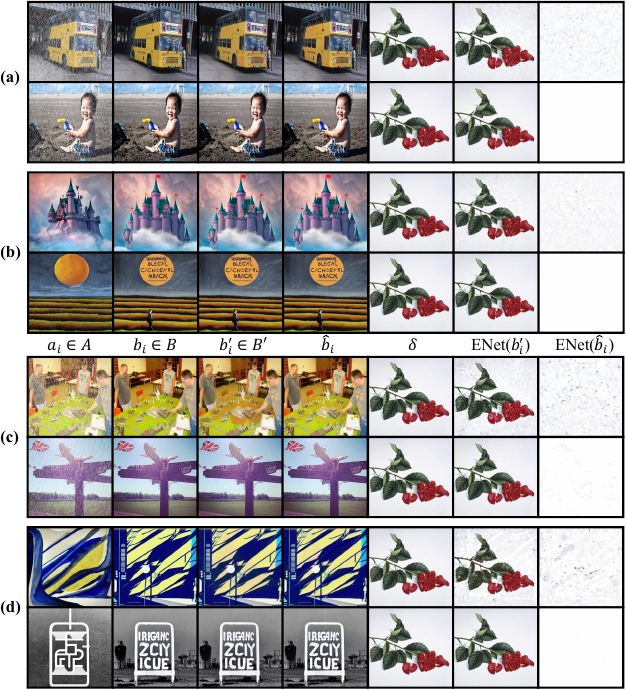}
  \caption{Qualitative demonstration of (a) $\text{HNet}^{-1}$ attack on the deraining task and (b) image generation task, with the first row attacking $\text{V}_\text{Wu}$ \cite{wu2020watermarking} and the second row attacking $\text{V}_\text{Zhang}$ \cite{zhang2024robust}. Similarly, (c) Forward HNet attack on the deraining task and (d) image generation task, with the first row attacking $\text{V}_\text{Wu}$ \cite{wu2020watermarking} and the second row attacking $\text{V}_\text{Zhang}$ \cite{zhang2024robust}.}
  \label{fig:qualitative_result}
\end{figure}

\subsubsection{Watermark Removal}
The qualitative results of our proposed attacks against $\text{V}_\text{Wu}$ \cite{wu2020watermarking} and  $\text{V}_\text{Zhang}$ \cite{zhang2024robust} are presented in Figure \ref{fig:qualitative_result}. In each sub-figure, the attack against deraining task is shown in the first row, while attack against image generation task is shown in the second row. Columns from left to right represent to-be processed image ($a_i \in A$), processed unmarked image ($b_i \in B$), processed marked image ($b_i' \in B'$), watermark removed image $\hat{b}_i \in \hat{B}$, embedded watermark ($\delta$), ENet extracted watermark from $b_i' \in B'$, and ENet extracted output from watermark removed $\hat{b}_i$. It can be seen in the figures that both proposed attacks can successfully remove the watermark, although dispersed noise dots remain when attacking $\text{V}_\text{Wu}$.

\begin{figure}[!t]
  \centering
  \includegraphics[width=1.0\columnwidth]{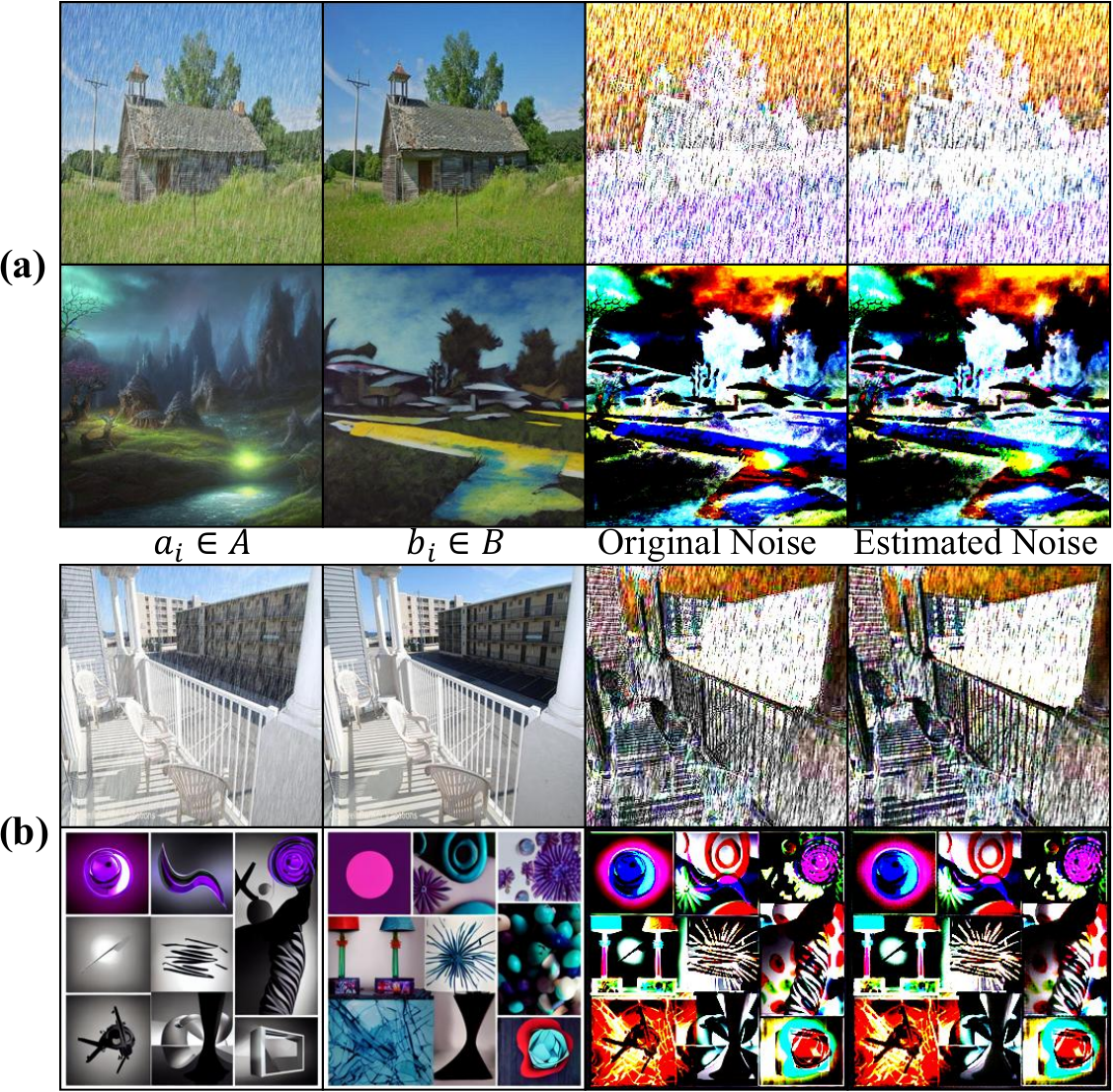}
  \caption{Qualitative demonstration of the proposed forward HNet attack on noise estimation performance for image deraining and generation tasks, where the noise is amplified by $10 \times$ for better visibility. (a) Against $\text{V}_\text{Wu}$ \cite{wu2020watermarking} (b) Against $\text{V}_\text{Zhang}$ \cite{zhang2024robust}.}
  \label{fig:noise_estimation}
\end{figure}

\subsubsection{Noise Estimation}
\label{sec:noise_estimation}
With a slight abuse of terms, we call the to-be-processed image component for GNet, i.e., $a_i-b_i$, collectively as ``noise'' (it is actually the rain component in deraining and transformation residual in image generation). Recall (\ref{eq:leakage4}), we note that the key factor enabling the \mbox{forward HNet} attack is the tractable estimation of the noise component $({\rm{SNet}}(a_i) - b_i')$.
Figure \ref{fig:noise_estimation} shows the qualitative experimental results of the proposed forward HNet attack for noise estimation in both deraining and image generation tasks against the two victim models. The columns from left to right represent the to-be-processed image ($a_i \in A$), processed unmarked image ($b_i \in B$), the ground-truth noise, and the estimated noise, respectively. Both noise images are amplified $10\times$ for better visibility. It can be seen that the estimated noise patterns show significantly high similarity to their respective ground-truth patterns, and the normalized correlation coefficients between the estimate and ground-truth noise patterns are $0.883$, $0.981$, $0.892$, and $0.979$, respectively, from top to bottom.

\begin{table*}[ht]
\centering
\renewcommand{\arraystretch}{1.1} 
\setlength{\tabcolsep}{8pt} 
\caption{Verification for approximation $\delta'_{ai} \approx \delta'_{bi}$ in forward HNet attack.}
\label{tab:verify_delta_a_b} 
\begin{tabular}{c|ccc|ccc}
\hline
\hline
\multirow{2}{*}{Tasks} & \multicolumn{3}{c|}{$\text{V}_{\text{Wu}}$ \cite{wu2020watermarking}}               & \multicolumn{3}{c}{$\text{V}_\text{Zhang}$ \cite{zhang2024robust}}  \\ \cline{2-7} 
   & PSNR (dB)$\uparrow$ & MS-SSIM$\uparrow$ & Correlation$\uparrow$ & PSNR (dB)$\uparrow$ & MS-SSIM$\uparrow$ & Correlation$\uparrow$\\ \hline
Deraining& $34.19$ & $0.948$ & $0.800$ & $34.41$ & $0.9513$ & $0.635$\\ \hline
Image Generation & $37.48$ & $0.987$ & $0.923$ & $35.84$ & $0.985$ & $0.894$\\ \hline
\hline
\end{tabular}
\end{table*}

\begin{table*}[ht]
    \fontsize{6}{6.5}\selectfont
    \centering
    \setlength{\tabcolsep}{1.5mm}
    \renewcommand{\arraystretch}{1.2}

    \caption{Quantitative evaluations and comparisons of proposed attacks against existing methods, where PSNR is in dB, and $0\leq \text{Correlation, }\text{MS-SSIM, }{\rm{SR}}_\text{Remove} \leq 1$.}
  \label{tab:quantitative_result} 
    
    \begin{tabular}{c|c|c|c|c|c|c|c|c|c}
    \hline
    \hline
    \multirow{2}{*}{\begin{tabular}[c]{@{}c@{}}Victim\\Model\end{tabular}} & 
    \multirow{2}{*}{\begin{tabular}[c]{@{}c@{}}Removal\\Attack\end{tabular}} &
    \multicolumn{4}{c|}{Deraining} & 
    \multicolumn{4}{c}{Image Generation} \\ \cline{3-10} 
    
     & & Correlation$\uparrow$ & PSNR (dB)$\uparrow$ & MS-SSIM$\uparrow$ & ${\rm{SR}}_\text{Remove}\uparrow$ & Correlation$\uparrow$ & PSNR (dB)$\uparrow$ & MS-SSIM$\uparrow$ & ${\rm{SR}}_\text{Remove}\uparrow$ \\ \hline

    \multirow{10}{*}{\makecell{$\text{V}_\text{Wu}$ \\ \cite{wu2020watermarking}}} & JPEG-20\% & \multirow{9}{*}{--} & $26.53$ & $0.940$ & $1.000$  &  \multirow{9}{*}{--} & $24.71$ & $0.936$ & $1.000$ \\ \cline{2-2} \cline{4-6}  \cline{8-10}   
    & JPEG-50\% & & $28.13$ & $0.960$ & $1.000$ & & $26.60$ & $0.957$ & $1.000$ \\ \cline{2-2} \cline{4-6}  \cline{8-10}  
    & AWGN-20dB & & $24.34$ & $0.906$ & $1.000$ & & $24.41$ & $0.931$ & $0.570$ \\ \cline{2-2} \cline{4-6}  \cline{8-10}  
    & AWGN-30dB & & $28.37$ & $0.958$ & $0.280$ & & $28.52$ & $0.967$ & $0.000$ \\ \cline{2-2} \cline{4-6}  \cline{8-10}  
    & Lattice-Interval2 \cite{liu2023erase} & & $13.82$ & $0.681$ & $1.000$ & & $13.73$ & $0.734$ & $1.000$ \\ \cline{2-2} \cline{4-6}  \cline{8-10}  
    & Lattice-Interval8 \cite{liu2023erase}& & $24.13$ & $0.915$ & $0.057$ & & $24.07$ & $0.935$ & $0.000$ \\ \cline{2-2} \cline{4-6}  \cline{8-10}  
    & WEvade-B-Q \cite{jiang2023evading}& & $35.98$ & $0.856$ & $0.380$ & & $43.11$ & $0.954$ & $0.140$ \\ \cline{2-2} \cline{4-6}  \cline{8-10}  
    & Regeneration-VAE \cite{zhao2024invisible} & & $32.89$ & $0.980$ & $1.000$ & & $31.18$ & $0.980$ & $0.970$ \\ \cline{2-2} \cline{4-6}  \cline{8-10}  
    & Regeneration-Diff \cite{zhao2024invisible}& & $22.74$ & $0.845$ & $1.000$ & & $21.38$ & $0.841$ & $1.000$ \\ \cline{2-2} \cline{4-6}  \cline{8-10}  
    & $\text{HNet}^{-1}$ \textbf{(Ours)} &  & $33.38$ & $0.987$ & $1.000$ &  & $31.75$ & $0.986$ & $1.000$\\ \cline{2-10} 
    & Forward HNet \textbf{(Ours)} & $0.883$ & $33.75$ & $0.990$ & $1.000$ & $0.981$ & $32.05$ & $0.988$ & $1.000$\\ \hline \hline

    \multirow{11}{*}{\makecell{$\text{V}_\text{Zhang}$ \\ \cite{zhang2024robust}}} & JPEG-20\% & \multirow{10}{*}{--} & $27.57$ & $0.955$ & $1.000$ & \multirow{10}{*}{--} & $24.93$ & $0.943$ & $1.000$ \\ \cline{2-2} \cline{4-6}  \cline{8-10}  
    & JPEG-50\% & & $29.86$ & $0.977$ & $1.000$ & & $26.78$ & $0.965$ & $1.000$ \\ \cline{2-2} \cline{4-6} \cline{8-10}    
    & AWGN-20dB & & $25.26$ & $0.925$ & $0.998$ & & $24.67$ & $0.943$ & $0.995$ \\ \cline{2-2} \cline{4-6}  \cline{8-10}    
    & AWGN-30dB & & $30.89$ & $0.980$ & $0.384$ & & $28.94$ & $0.980$ & $0.010$ \\ \cline{2-2} \cline{4-6} \cline{8-10}   
    & Lattice-Interval2 \cite{liu2023erase}& & $13.89$ & $0.695$ & $1.000$ & & $13.75$ & $0.745$ & $1.000$ \\ \cline{2-2} \cline{4-6} \cline{8-10}     
    & Lattice-Interval8 \cite{liu2023erase}& & $24.94$& $0.956$ & $0.384$ & & $24.30$ & $0.947$ & $0.950$ \\ \cline{2-2} \cline{4-6} \cline{8-10}    
    & WEvade-B-Q \cite{jiang2023evading}& & $30.95$ & $0.796$ & $0.530$ & & $35.00$ & $0.884$ & $0.450$ \\ \cline{2-2} \cline{4-6} \cline{8-10}   
    & Regeneration-VAE \cite{zhao2024invisible}& & $32.67$ & $0.978$ & $1.000$ & & $32.54$ & $0.986$ & $0.980$ \\ \cline{2-2} \cline{4-6} \cline{8-10}
    & Regeneration-Diff \cite{zhao2024invisible}& & $22.70$ & $0.858$ & $1.000$ & & $21.61$ & $0.853$ & $1.000$ \\ \cline{2-2} \cline{4-6} \cline{8-10}  
    
    & $\text{HNet}^{-1}$ \textbf{(Ours)} &  & $33.98$ & $0.988$ & $1.000$&  & $31.41$ & $0.991$ & $1.000$ \\ \cline{2-10} 
    & Forward HNet \textbf{(Ours)} & $0.892$ & $34.69$ & $0.992$ & $1.000$ & $0.979$ & $32.60$ & $0.990$ & $1.000$\\ \hline
    \hline
    \end{tabular}
    
\end{table*}

\subsection{Quantitative Results}
\label{sec:quantitative_result}
The verification results for $\delta'_{ai} \approx \delta'_{bi}$ are shown in Table~\ref{tab:verify_delta_a_b}, while
the quantitative experimental results applying our proposed attacks against the two victim models for deraining and image generation are summarized in Table \ref{tab:quantitative_result}, where the Correlation column refers to the average normalized correlation coefficients between the estimated and ground-truth noises for the proposed forward HNet attack.

\subsubsection{Verification for $\delta'_{ai} \approx \delta'_{bi}$} 
\label{sec:verification_ai_bi}

Table \ref{tab:verify_delta_a_b} presents the average PSNR, MS-SSIM, and normalized correlation coefficient between $\delta'_{ai}$ and $\delta'_{bi}$. The consistently high values of these metrics strongly support the hypothesis $\delta'_{ai} \approx \delta'_{bi}$ in the derivation of forward HNet attack.

\subsubsection{Fidelity}
The fidelity of the proposed attacks is evaluated using PSNR and MS-SSIM metrics, comparing $\hat{b}_i$ with the unknown ground truth $b_i$. As shown in Table \ref{tab:quantitative_result}, for the $\text{HNet}^{-1}$ attack, the average PSNR values exceed $33.38$ dB for deraining task and $31.41$ dB for image generation task, with all MS-SSIM values surpassing $0.986$.
For the forward HNet attack, the average PSNR values are greater than $33.75$ dB for deraining and $32.05$ dB for image generation, while all MS-SSIM values are greater than $0.988$, for both victims models, demonstrating very high-fidelity performance. The fidelity comparison of the two attacks demonstrate the superior image quality performance achieved by the forward HNet attack. In addition, we notice that the image quality of the generated image data is inferior to that of the deraining data. As illustrated in Figure \ref{fig:qualitative_result}, generated images are significantly more complex than those in the PASCAL VOC dataset \cite{everingham2010pascal}, resulting in the task-wise image quality differences.

\subsubsection{Removal Success Rate}
All eight sets of experimental results shown in Table \ref{tab:quantitative_result} have achieved $100\%$ watermark removal (${\rm{SR}}_\text{Remove} = 1.000$). This represents a significant advancement over prior watermark removal methods not specifically designed for box-free model watermarking, as our attacks maintain the perfect removal rate while simultaneously preserving superior image quality.

\subsection{Ablation Study}
\begin{figure}[!t]
  \centering
  \includegraphics[width=1.0\columnwidth]{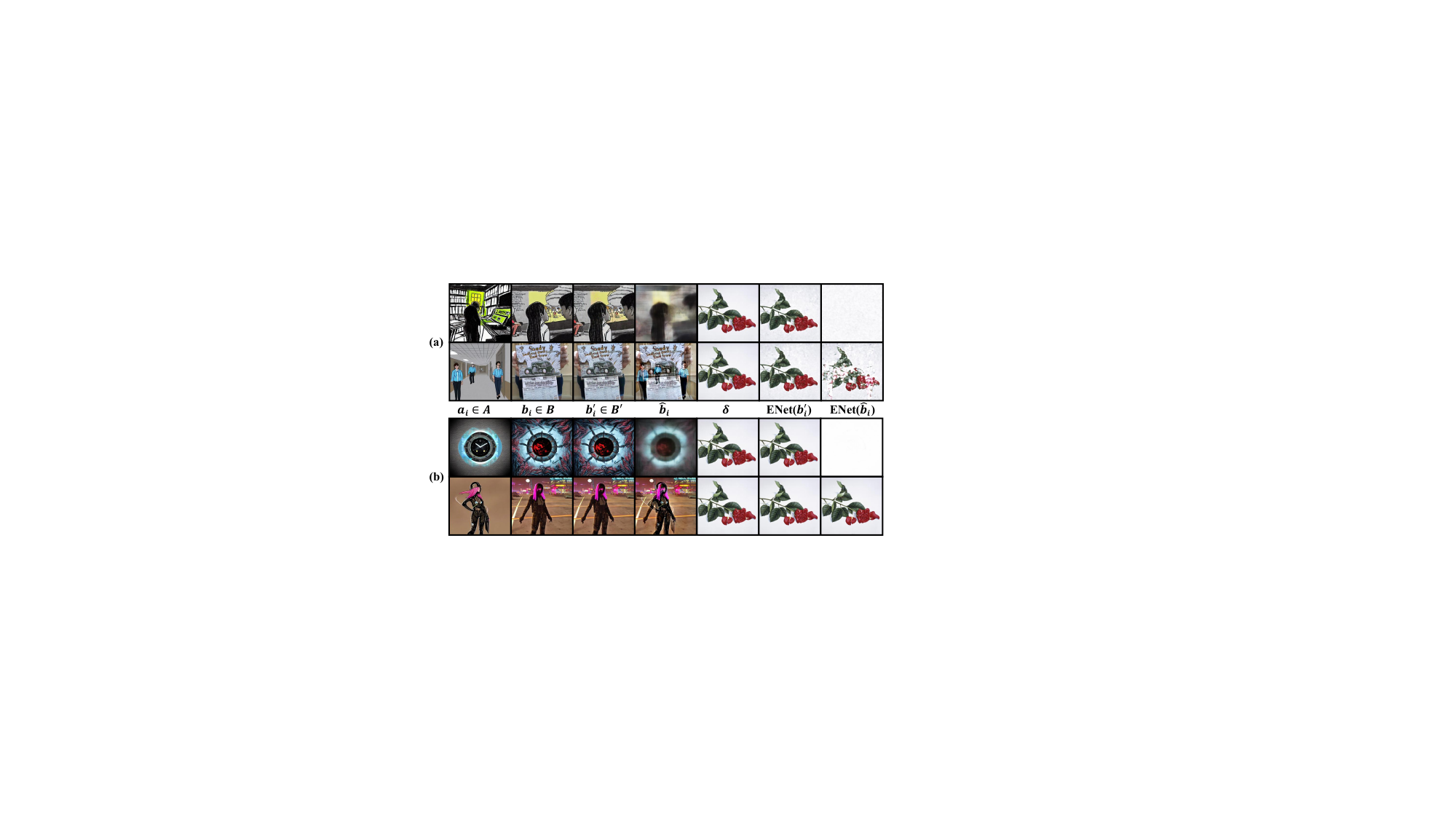}
  \caption{Ablation study on the need for bypassing GNet in the proposed attack for image generation task, against (a) $\text{V}_\text{Wu}$ \cite{wu2020watermarking} and (b) $\text{V}_\text{Zhang}$ \cite{zhang2024robust}. In each subplot, the first row is $\text{HNet}^{-1}$ attack result, while the second is forward $\text{HNet}$ attack result.}
  \label{fig:ablation_study}
  \vspace*{-6pt}
\end{figure}

In (\ref{eq:bypass_gnet}), curated images that satisfy approximate identity transformation are used as input to bypass GNet encapsulated in ONet, which are subsequently combined with the corresponding watermarked images to conduct attacks. We demonstrate the necessity of bypassing GNet in the ablation study where image generation task is considered as an example. New $\text{HNet}^{-1}$ and SNet are trained using pairs of images in $A$ and their corresponding watermarked images in $B'$. The attack results are illustrated in Figure \ref{fig:ablation_study}. We notice that for the two victims and the two attacks, $\text{HNet}^{-1}$ (the first row of each subfigure) removes the embedded watermark (the last column) but at the cost of significant image quality degradation (the fourth column), thereby failing to meet the fidelity requirement of our attack goal. In addition, forward HNet (the second row of each subfigure) fails to remove the embedded watermark (the last column), while incorporating the original components from $a_i$ into $\hat{b}_i$. These results collectively demonstrate the need for circumventing GNet for conducting successful attacks. 

\section{Conclusion}
We have proposed two black-box watermark removal attacks driven by query-based reverse engineering against existing box-free model watermarking under the real-world black-box setting, specifically image-to-image models. We begin by showing the simple and efficient attack which removes watermarks by training an inversion model of HNet but with inferior performance on output image quality. To fill the gap, we demonstrate
the equivalent additive form of box-free model watermarking originally performed by the nonlinear HNet and exploit a simple surrogate training under the practical threat model. Then, a forward HNet attack is developed, which, based on the surrogate model of HNet, the query input, and the ONet output, can effectively estimate the unknown processed and unmarked image, thus achieving watermark removal with better image quality preservation. Finally, we point out the necessity of API detection in box-free model watermarking system against these attacks. Extensive experiments against the victim models on deraining and image generation tasks demonstrated that our proposed attacks can remove embedded watermarks at perfect success rates. Overall, our proposed removal attack reveals the vulnerabilities of box-free model watermarking in real-world scenarios, highlighting the urgent need for more effective countermeasures.

\newpage
\bibliography{aaai2026}

\begin{thebibliography}{15}
\providecommand{\natexlab}[1]{#1}

\bibitem[{An et~al.(2025)An, Hua, Fang, Xu, Rahardja, and Fang}]{an2025decoder}
An, H.; Hua, G.; Fang, Z.; Xu, G.; Rahardja, S.; and Fang, Y. 2025.
\newblock Decoder Gradient Shield: Provable and High-Fidelity Prevention of Gradient-Based Box-Free Watermark Removal.
\newblock In \emph{Proceedings of the Computer Vision and Pattern Recognition Conference}, 13424--13433.

\bibitem[{Everingham et~al.(2010)Everingham, Van~Gool, Williams, Winn, and Zisserman}]{everingham2010pascal}
Everingham, M.; Van~Gool, L.; Williams, C.~K.; Winn, J.; and Zisserman, A. 2010.
\newblock The pascal visual object classes (voc) challenge.
\newblock \emph{International journal of computer vision}, 88: 303--338.

\bibitem[{Huang et~al.(2023)Huang, Li, Cai, Wang, Guo, Fang, Chen, and Wang}]{huang2023can}
Huang, Z.; Li, B.; Cai, Y.; Wang, R.; Guo, S.; Fang, L.; Chen, J.; and Wang, L. 2023.
\newblock What can discriminator do? towards box-free ownership verification of generative adversarial networks.
\newblock In \emph{Proceedings of the IEEE/CVF international conference on computer vision}, 5009--5019.

\bibitem[{Jiang, Zhang, and Gong(2023)}]{jiang2023evading}
Jiang, Z.; Zhang, J.; and Gong, N.~Z. 2023.
\newblock Evading watermark based detection of AI-generated content.
\newblock In \emph{Proceedings of the 2023 ACM SIGSAC Conference on Computer and Communications Security}, 1168--1181.

\bibitem[{Johnson, Alahi, and Fei-Fei(2016)}]{johnson2016perceptual}
Johnson, J.; Alahi, A.; and Fei-Fei, L. 2016.
\newblock Perceptual losses for real-time style transfer and super-resolution.
\newblock In \emph{Computer Vision--ECCV 2016: 14th European Conference, Amsterdam, The Netherlands, October 11-14, 2016, Proceedings, Part II 14}, 694--711. Springer.

\bibitem[{Liu et~al.(2023{\natexlab{a}})Liu, Xiang, Guo, Li, Zhang, and Liao}]{liu2023erase}
Liu, H.; Xiang, T.; Guo, S.; Li, H.; Zhang, T.; and Liao, X. 2023{\natexlab{a}}.
\newblock Erase and Repair: An Efficient Box-Free Removal Attack on High-Capacity Deep Hiding.
\newblock \emph{IEEE Transactions on Information Forensics and Security}, 18: 5229--5242.

\bibitem[{Liu et~al.(2023{\natexlab{b}})Liu, Han, Ma, and et~al.}]{LIU2023100017}
Liu, Y.; Han, T.; Ma, S.; and et~al. 2023{\natexlab{b}}.
\newblock Summary of ChatGPT-Related research and perspective towards the future of large language models.
\newblock \emph{Meta-Radiology}, 1(2): 100017.

\bibitem[{Rombach et~al.(2022)Rombach, Blattmann, Lorenz, Esser, and Ommer}]{rombach2022high_stable_diffusion}
Rombach, R.; Blattmann, A.; Lorenz, D.; Esser, P.; and Ommer, B. 2022.
\newblock High-resolution image synthesis with latent diffusion models.
\newblock In \emph{Proceedings of the IEEE/CVF conference on computer vision and pattern recognition}, 10684--10695.

\bibitem[{Wang, Simoncelli, and Bovik(2003)}]{wang2003multiscale}
Wang, Z.; Simoncelli, E.; and Bovik, A. 2003.
\newblock Multiscale structural similarity for image quality assessment.
\newblock In \emph{The Thrity-Seventh Asilomar Conference on Signals, Systems {\&} Computers}, volume~2, 1398--1402.

\bibitem[{Wu et~al.(2020)Wu, Liu, Yao, and Zhang}]{wu2020watermarking}
Wu, H.; Liu, G.; Yao, Y.; and Zhang, X. 2020.
\newblock Watermarking neural networks with watermarked images.
\newblock \emph{IEEE Transactions on Circuits and Systems for Video Technology}, 31(7): 2591--2601.

\bibitem[{Zhang and Patel(2018)}]{zhang2018density}
Zhang, H.; and Patel, V.~M. 2018.
\newblock Density-aware single image de-raining using a multi-stream dense network.
\newblock In \emph{Proceedings of the IEEE conference on computer vision and pattern recognition}, 695--704.

\bibitem[{Zhang et~al.(2020)Zhang, Chen, Liao, Fang, Zhang, Zhou, Cui, and Yu}]{zhang2020model}
Zhang, J.; Chen, D.; Liao, J.; Fang, H.; Zhang, W.; Zhou, W.; Cui, H.; and Yu, N. 2020.
\newblock Model watermarking for image processing networks.
\newblock In \emph{Proceedings of the AAAI conference on artificial intelligence}, volume~34, 12805--12812.

\bibitem[{Zhang et~al.(2024)Zhang, Chen, Liao, Ma, Fang, Zhang, Feng, Hua, and Yu}]{zhang2024robust}
Zhang, J.; Chen, D.; Liao, J.; Ma, Z.; Fang, H.; Zhang, W.; Feng, H.; Hua, G.; and Yu, N. 2024.
\newblock Robust Model Watermarking for Image Processing Networks via Structure Consistency.
\newblock \emph{IEEE Transactions on Pattern Analysis and Machine Intelligence}, 1--8.

\bibitem[{Zhang et~al.(2021)Zhang, Chen, Liao, Zhang, Feng, Hua, and Yu}]{zhang2021deep}
Zhang, J.; Chen, D.; Liao, J.; Zhang, W.; Feng, H.; Hua, G.; and Yu, N. 2021.
\newblock Deep model intellectual property protection via deep watermarking.
\newblock \emph{IEEE Transactions on Pattern Analysis and Machine Intelligence}, 44(8): 4005--4020.

\bibitem[{Zhao et~al.(2024)Zhao, Zhang, Su, Vasan, Grishchenko, Kruegel, Vigna, Wang, and Li}]{zhao2024invisible}
Zhao, X.; Zhang, K.; Su, Z.; Vasan, S.; Grishchenko, I.; Kruegel, C.; Vigna, G.; Wang, Y.-X.; and Li, L. 2024.
\newblock Invisible image watermarks are provably removable using generative ai.
\newblock \emph{Advances in Neural Information Processing Systems}, 37: 8643--8672.

\end{thebibliography}

\newpage

\end{document}